\documentclass[epj]{svjour}
\usepackage{graphics}
\begin{document}
\title{Klein tunneling of two correlated bosons}
\author{Stefano Longhi \inst{1} \and Giuseppe Della Valle \inst{1}
}                     
\offprints{}          
\institute{Dipartimento di Fisica, Politecnico di Milano and Istituto di Fotonica e Nanotecnologie del Consiglio Nazionale delle Ricerche, Piazza Leonardo da Vinci 32, I-20133 Milano (Italy)}
\date{Received: date / Revised version: date}
%
\abstract{Reflection of two strongly interacting bosons with long-rage interaction hopping on a one-dimensional lattice scattered off by a potential step is theoretically  investigated in the framework of the extended Hubbard model.  The analysis shows that, in the presence of unbalanced on-site and nearest-neighbor site interaction,  two strongly correlated bosons forming a bound particle state can penetrate a high barrier, despite the single particle can not. Such a phenomenon is analogous to one-dimensional Klein tunneling of a relativistic massive Dirac particle across a potential step.
} 
\maketitle

\section{Introduction}

One of the most intriguing predictions of relativistic quantum mechanics
is that a below-barrier electron can pass a large repulsive and sharp
potential step (of the order of twice the rest energy $mc^2$
of the electron) without the exponential damping expected
for a nonrelativistic particle. Such a transparency effect,
originally predicted by Klein \cite{Klein} and referred to as Klein
tunneling (KT) \cite{Calogero}, is related to the existence of negative-energy
states of the Dirac equation.  The observation of KT for a relativistic particle is very challenging, because it would require an ultrastrong field, of the order of the critical  field for $e^-e^+$ pair production in vacuum \cite{Calogero,Sauter}, which is not currently available. In recent years, there has been an increased interest in
simulating KT in diverse and experimentally accessible  physical
 systems (see, for instance, \cite{K1,K2,K3,K4,K5,K6,K7,K8,K9,K10,K11,K12,K13,K14,K15,K16,K17,K18,K19} and references therein).
A remarkable example is provided by electronic transport in graphene, 
a carbon mono
layer of honeycomb shape, where the energy dispersion relation
near a Dirac point resembles the
dispersion of relativistic electrons \cite{note}. 
Experimental evidences
for KT have been reported in graphene heterojunctions
\cite{K3}, carbon nanotubes \cite{K5}, cold ions in
Paul traps \cite{K16}, cold atoms in optical lattices \cite{K17}, and photonic superlattices \cite{K19}.\par 
Such previous studies have 
been mainly devoted  to the simulation of KT of non-interacting particles, whereas less attention has been paid to the role of particle interaction. In Ref.\cite{K8} it was shown that 
KT of relativistic electrons in graphene is strongly suppressed taking into account electron-electron interaction. 
In this work we show, conversely, that in the framework of an extended Bose-Hubbard model two strongly-interacting bosons hopping on a one-dimensional lattice and scattered off by a potential step 
can show a tunneling effect that resembles KT of a relativistic massive Dirac particle, i.e. they can be partially transmitted across a sufficiently high potential barrier, despite a single particle can not.  Such a correlation-induced KT  
is associated to the formation of a bound (molecular) particle state \cite{uffi1,uffi2,uffi3,uffi3b,uffi4,uffi5,uffi6,uffi7}, which behaves differently from the single particle state as it is scattered off by a potential barrier \cite{uffi7,note2} or when an external field is applied \cite{uffi4,uffi5,uffi8a,uffi8b,uffi9,uffi10}. 
We emphasize that, for the observation of correlation-induced KT, it is crucial that the particles exhibit long-range (nearest-neighbor) interaction, with the existence of two minibands for the two-particle bound state. Once a potential step is applied to the lattice, tunneling between the two minibands, which is formally analogous to one-dimensional KT of a massive Dirac particle, can occur. Unlike one-dimensional KT of a single-particle in a superlattice previously investigated in Refs.\cite{K12,K19}, in the Hubbard model the potential step is impenetrable for the single particle, and KT is  a clear signature of long-range {\it particle interaction}.  We will also show that KT of a bound particle state can be observed even in the absence of nearest-neighbor particle interaction, i.e. in the framework of a standard Bose-Hubbard model with on-site particle interaction solely, provided that an external high-frequency ac driving force is applied.\par
The paper is organized as follows. In Sec.2 the tunneling dynamics of two strongly-correlated particles scattered off by a potential step is investigated in the framework of an extended Bose-Hubbard model. The analysis clearly shows that, while a below-barrier {\it single} particle is fully reflected from the potential step,   a two-particle bound state can penetrate into the barrier owing to an interband tunneling process which is fully analogous to relativistic one-dimensional KT of a massive particle. In Sec.3 we consider the tunneling  dynamics of a bound particle state in the framework of a standard Bose-Hubbard model,  i.e. without nearest-neighbor particle interaction, and show that KT can be observed as well by application of an external high-frequency driving force. The predictions of the theoretical analysis and the onset of KT for a two-particle bound state are confirmed in Sec.4 by numerical simulations of both the extended Bose-Hubbard model and the ac-driven Bose-Hubbard model in the two-particle sector of Fock space. Finally, the main conclusions are outlined in Sec.5, including a brief discussion on a possible observation of the predicted phenomenon in a model system of the two-particle Bose-Hubbard model. 

\section{Klein tunneling of a two-particle bound state in the extended Bose-Hubbard model}

\subsection{The model}
We consider the hopping dynamics of two strongly-interacting particles on a tight-binding one-dimensional lattice in the presence of a potential barrier with both on-site and nearest-neighbor interaction. The two particles can be  two bosons, such as two neutral atoms trapped on a one-dimensional lattice, or two fermions, such as two electrons with opposite spins. For the sake of definiteness, we will refer to the former case. The particle dynamics can be described by a rather standard 
 one-dimensional extended Bose-Hubbard
model (EHM) \cite{uffi6,uffi7,uffi8a,uffi8b} with Hamiltonian ($\hbar=1$)
\begin{eqnarray}
\hat{H} & = & - J \sum_{l} \hat{a}^{\dag}_{l} \left( \hat{a}_{l-1}+\hat{a}_{l+1}  \right) +\frac{U}{2} \sum_{l} \hat{n}_{l} (\hat{n}_{l}-1) \nonumber \\
&+ & V \sum_l \hat{n}_{l} \hat{n}_{l+1}+\sum_l \epsilon_l \hat{n}_{l}.
\end{eqnarray}
In Eq.(1) $\hat{a}^{\dag}_{l}$ are $\hat{a}_{l}$ are the creation and annihilation operators of bosons and  $\hat{n}_{l}=\hat{a}^{\dag}_{l} \hat{a}_{l}$ the particle number operators at lattice sites $l=0, \pm 1, \pm 2,...$, $J$ is the single-particle hopping rate between adjacent sites,  $U$ and $V$ define the
on-site and nearest-neighbor interaction energies, respectively, and $\epsilon_l$ is the applied potential step of height $\Delta$, defined by
\begin{equation}
\epsilon_l= \left\{  
\begin{array}{cl}
0 & l<0 \\
\Delta & l \geq 0.
\end{array}
\right.
\end{equation}
The Hamiltonian (1) conserves the total number $N$ of particles. As compared to the standard Hubbard or Bose-Hubbard model [which is obtained by letting $V=0$  in Eq.(1)], the EHM accounts for nonlocal particle interaction, which is essential for the observation of KT, as discussed below.
The EHM is a prototype model in condensed-matter physics \cite{EHMa,EHMb,EHMc,EHMd,EHMe,EHMf,EHMg}, where the nearest-neighbor term $V$ arises from  Coulomb repulsion of electrons in adjacent sites due to non-perfect screening of electronic
charges. Nearest-neighbor particle interaction also arise for fermionic ultracold atoms or molecules
with magnetic or electric dipole-dipole interactions in optical lattices. 
In this case the ratio $V/U$ can be
tuned by modifying
the trap geometry of the condensate, additional external dc
electric fields, combinations with fast rotating external fields,
etc. (see, for instance, \cite{DMa,DMb} and references therein). In the following we will consider the case $U,V>0$, corresponding to particle repulsion, and $U>V$ for the sake of definiteness. However, a similar analysis could be done for the attractive particle case $U,V <0$.\\ 

\subsection{Single-particle tunneling}

Tunneling  of a single particle on a tight-binding lattice scattered off by a potential step or a potential barrier is a rather simple problem, which has been studied in previous papers (see, for instance, \cite{scatt1,scatt2}).  The problem is here briefly reviewed for the sake of completeness. 
In the $N=1$ sector of Fock space, the state vector $|\psi(t) \rangle$ of the system can be expanded as $|\psi(t) \rangle = \sum_l c_l(t) \hat{a}^{\dag}_l | 0 \rangle$. The amplitude probabilities $c_l(t)$ to find the particle at lattice site $l$ evolve according to the coupled equations
\begin{equation}
i \frac{dc_l}{dt}=-J(c_{l+1}+c_{l-1})+\epsilon_l c_l.
\end{equation}
In the absence of the potential step ($\epsilon_l=0$), the single-particle Bloch eigenstates of the system are plane waves $c_l(t) \propto \exp(iql) \exp[-i E(q) t]$, where $ -\pi \leq q < \pi$ is the quasi-momentum and $E(q)=-2 J \cos (q)$ is the dispersion relation of the tight-binding band [see Fig.1(a)]. A particle wave-packet with a carrier quasi-momentum $q=q_0$ and mean energy $E_0=-2 J \cos (q_0)$ moves on the lattice with a group velocity $v_g= (dE/dq)_{q_0}=2J \sin (q_0)$.  In the presence of the potential step of height $\Delta$ [see Eq.(2)], the space-dependent band diagram of the particle is schematically shown in Figs.1(b) and (c) for increasing values of the potential height $\Delta$. For a relatively low barrier height, under-barrier transmission occurs, with the wave packet partially transmitted and partially reflected from the potential step [see Fig.1(b)]; however, as $\Delta$ is increased such that  $\Delta >2J+E_0$, the potential step is impenetrable, and the wave packet is fully reflected: the barrier step is impenetrable for a single particle [see Fig.1(c)]; in particular, for $\Delta > 4J$ any particle is fully reflected from the potential step. Note that this tunneling scenario is analogous to that of a single non-relativistic particle freely moving and scattered off by a potential step (i.e. without the periodic lattice  potential). In particular, partial (or full) transparency of the step as the barrier height $\Delta$ is increased, i.e. KT, can not be observed for a single particle.

\begin{figure}
\resizebox{0.45\textwidth}{!}{
  \includegraphics{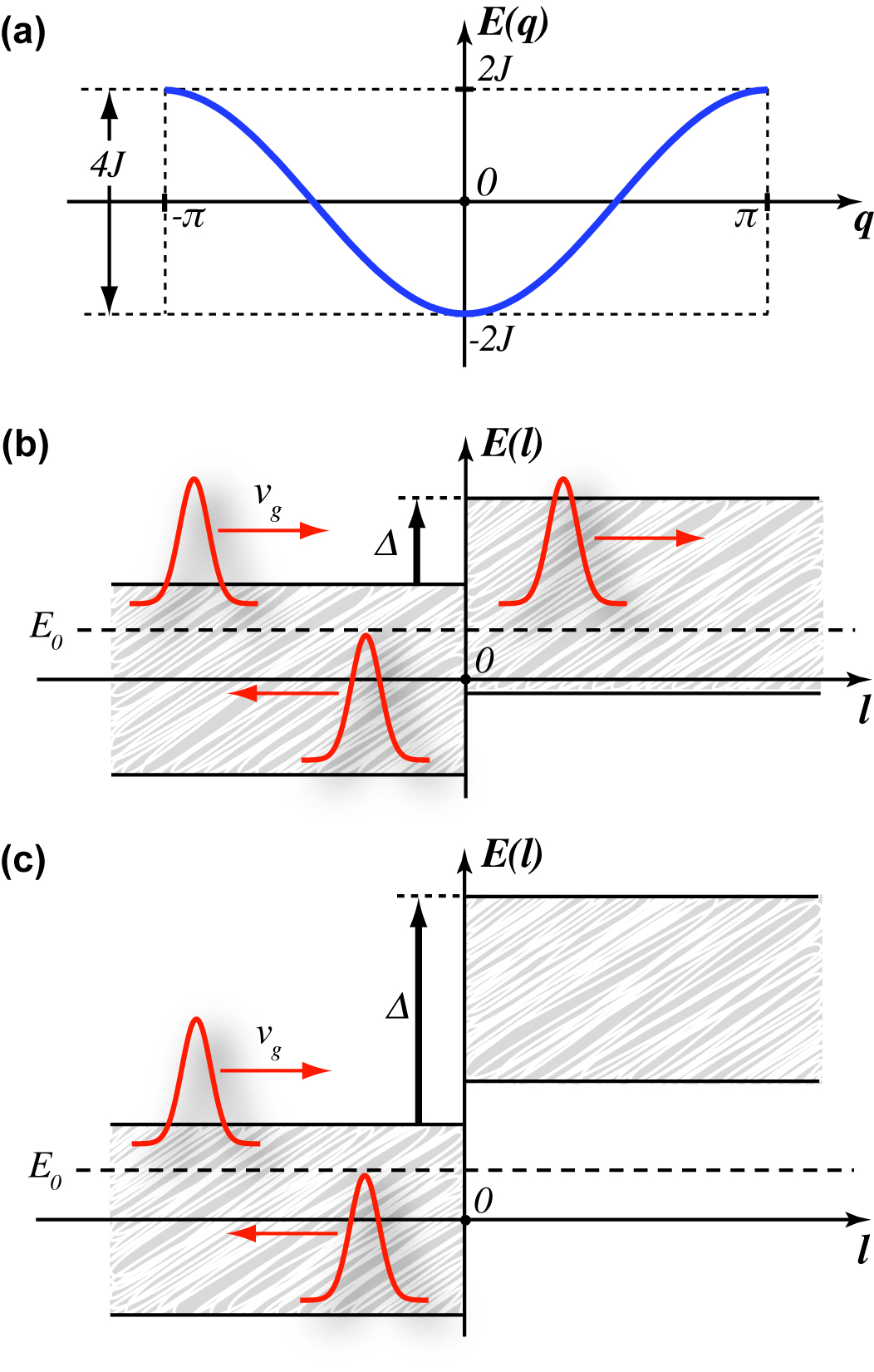}}
\caption{Tunneling of a single-particle hopping on a one-dimensional tight-binding lattice scattered off by a potential step. (a) Band diagram of the particle. (b) and (c): Space-dependent band diagram and wave packet scattering in the presence of a potential step. In (b) the above-barrier tunneling is shown: a particle wave packet is partially reflected and partially transmitted by the potential step. The case of  under-barrier tunneling is shown in (c): the potential step is impenetrable and a particle wave packet is fully reflected.}
\end{figure}

\subsection{Two-particle tunneling}
To study the scattering properties of the potential step for a bound pair, let us consider the $N=2$ particle sector of the Fock space for the Hamiltonian (1) and let us expand the state vector $| \psi(t) \rangle$ of the system as 
 \begin{equation}
 | \psi(t) \rangle= \sum_{n,m}c_{n,m}(t) \hat{a}_{n}^{\dag} \hat{a}_{m}^{\dag}|0 \rangle,
 \end{equation}
 where $c_{n,m}(t)$ is the amplitude probability to find one particle at the lattice site $n$ and the other particle at the lattice site $m$, with $c_{n,m}=c_{m,n}$ for bosonic particles. 
  The evolution equations for the amplitude probabilities $c_{n,m}$, as obtained from the Schr\"{o}dinger equation $i \partial_t | \psi \rangle = \hat{H} | \psi \rangle$ with $\hbar=1$, read explicitly
 \begin{eqnarray}
 i \frac{dc_{n,m}}{dt} & = & - J  \left( c_{n+1,m}+c_{n-1,m}+c_{n,m-1}
  + c_{n,m+1} \right) \\
  & + & \left[ U \delta_{n,m} +V \delta_{n,m+1}+V \delta_{n,m-1}+(\epsilon_n+\epsilon_m) \right] c_{n,m}. \nonumber
   \end{eqnarray}
 Here we consider the strong interaction and low-field regimes, corresponding to $J, \Delta  \ll U,V$, with $U-V$ of the order of the tunneling rate $J$. In this regime at leading order the dynamics in Fock space for the amplitudes $c_{n,m}$ with $m=n, n \pm 1$ decouples from the other states (see, for instance, \cite{uffi9}). Therefore, if we assume that the two particles are initially placed at the same lattice site or in nearest sites, i.e. if we assume $c_{n,m}(0)=0$ for $m \neq n, n \pm 1$ as an initial condition, Eqs.(5) can be restricted to the amplitudes $c_{n,n}$ and $c_{n,n+1}=c_{n+1,n}$: the two particles form a bound state, and are allowed to stay solely at the same site or in nearest-neighbor sites owing to energy conservation constraints. After setting
\begin{eqnarray}
c_{n,n}(t) & = & f_{2n}(t) \exp[-i(U+V)t/2]  \\
c_{n,n+1}(t) & = & \frac{1}{\sqrt{2}}f_{2n+1}(t) \exp[-i(U+V)t/2]
\end{eqnarray}
the dynamics of the amplitudes $f_n(t)$ reads 
\begin{equation}
i \frac{d f_n}{dt}  = - \sqrt{2} J (f_{n+1}+f_{n-1})+(-1)^n \sigma f_n + \delta_n f_n
\end{equation}

\begin{figure*}
\resizebox{0.85\textwidth}{!}{
  \includegraphics{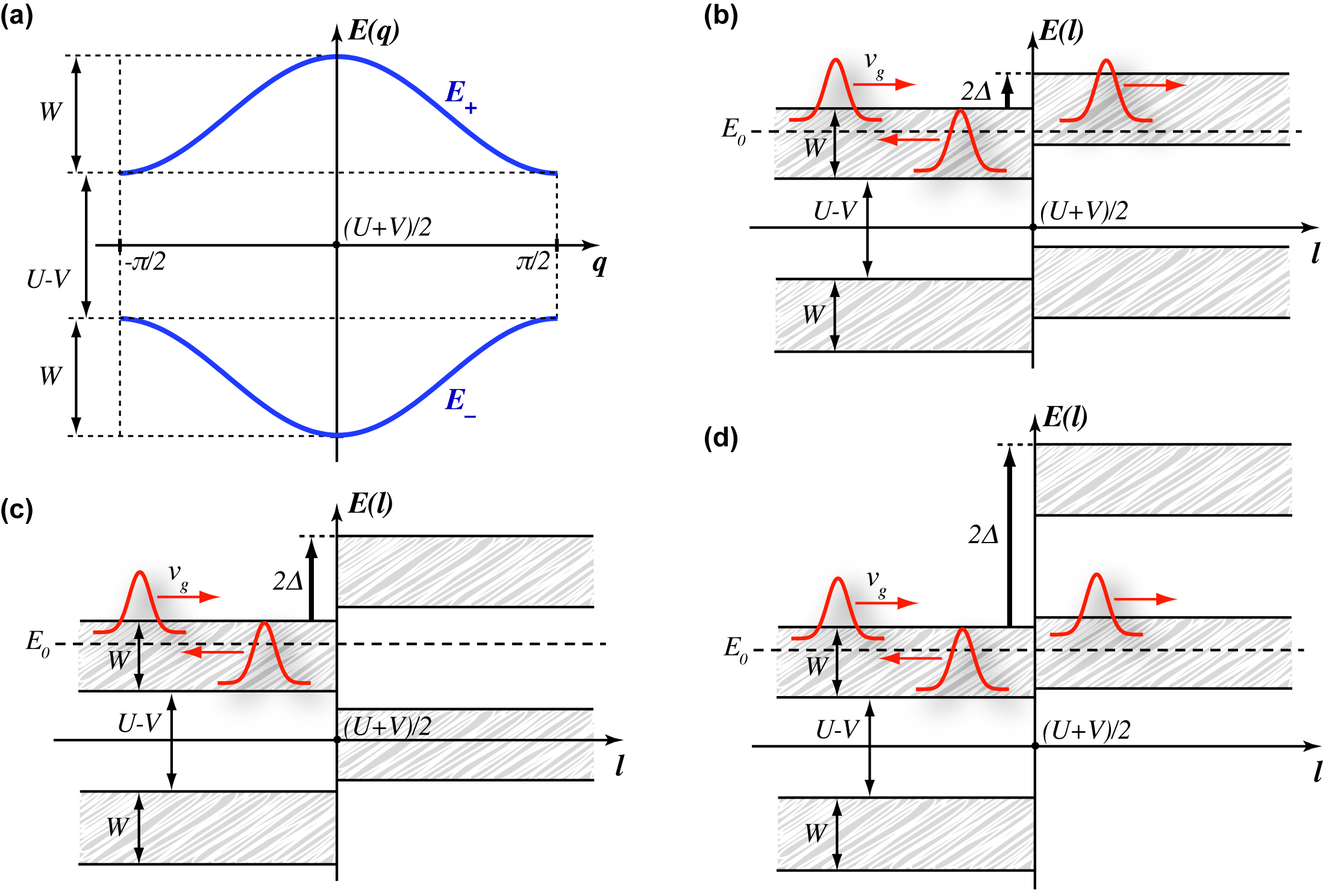}}
\caption{Tunneling of a bound-particle state in the framework of the extended Bose-Hubbard model.  (a) Band diagram of a two-particle bound state for $U>V$, showing the appearance of two minibands. (b), (c) and (d): Space-dependent band diagram and wave packet scattering in the presence of a potential step with increasing height $\Delta$. In (b) a two-particle bound wave packet is partially reflected and partially transmitted by the potential step. In (c)  the potential step is impenetrable and the particle wave packet is fully reflected. In (d) an interband tunneling process, with partial transmission of the wave packet, is observed. Such an interband two-particle tunneling process is analogous to KT of a massive Dirac particle.}
\end{figure*}

where we have set 
\begin{equation}
\sigma \equiv \frac{U-V}{2}
\end{equation}
and
\begin{equation}
\delta_n \equiv  \left\{  
\begin{array}{cl}
2 \epsilon_{n/2} & n \; {\rm even} \\
\epsilon_{(n+1)/2}+\epsilon_{(n-1)/2} & n \; {\rm odd}.
\end{array}
\right.
\end{equation}
Note that for a potential step [Eq.(2)] one has explicitly
\begin{equation}
\delta_n \equiv  \left\{  
\begin{array}{cl}
0 & n  \leq -2 \\
\Delta & n=-1 \\
2 \Delta & n \geq 0 
\end{array}
\right.
\end{equation}
Equations (8) are formally analogous to the tight-binding model describing the hopping dynamics of a single particle in a binary superlattice in the presence of a potential barrier, described by Eq.(10), the energy difference  $2 \sigma$ between adjacent lattice sites being determined by the unbalance $(U-V)$ of on-site and nearest-neighbor site interaction in the original problem.  As discussed in several previous works (see, for instance, \cite{K12,K15,K19,Cannata,LonghiOL}),  a Dirac-like behavior is found for a non-relativistic particle hopping on a binary superlattice in one dimension, including the analogue of KT in the presence of a potential barrier.  In our case, since Eqs.(8) describe the correlated hopping dynamics of a particle bound state, tunneling of two correlated bosons across the potential barrier is thus expected to be analogous to relativistic KT.  To clarify this point, let us first observe that, in the absence of the barrier step ($\delta_n=0$) and for $U \neq V$, the two-particle bound state of the EHM is described by {\it two} minibands with the dispersion relations (see, e.g., \cite{uffi6,uffi7,uffi9})
 \begin{equation}
 E_{\pm}(q)  =  \frac{U+V}{2} \pm  \sqrt{\left( \frac{U-V}{2} \right)^2+8 J^2 \cos^2 (q)  }
 \end{equation}
and the corresponding Bloch eigenstates are given by
\begin{equation}
f_n \propto \left( 
\begin{array}{c}
-2 \sqrt{2} J \cos (q) \\
E(q)-\sigma-\frac{U+V}{2}
\end{array}
\right) \exp[iqn-iE(q)t]
\end{equation}
with $E(q)=E_{+}(q)$ or $E_{-}(q)$ for the two minibands [see Fig.1(b)]. In Eq.(13), the upper (lower) row applies to an even (odd) index $n$. The two bands are separated by the gap $2 \sigma=U-V$, and their width is given by
\begin{equation}
W=\sqrt{\left( \frac{U-V}{2}\right)^2+8J^2}-\left( \frac{U-V}{2}\right).
\end{equation}
Note that, for $\sigma \gg J$, the Bloch states of the upper ($E=E_+$) miniband basically correspond to occupation of the even-index sites, i.e. $f_n \simeq 0$ for $n$ odd. According to Eqs.(6) and (7), such states correspond to the two particles occupying the same lattice site. Conversely, the lower miniband  ($E=E_-$) corresponds to occupation of odd-index sites, i.e. $f_n \simeq 0$ for $n$ even. This means that, in this case, the two particles occupy nearest-neighbor sites [according to Eqs.(6) and (7)].  A wave packet with carrier wave number $q=q_0$, obtained as a superposition of Bloch states with wave number $q$ close to $q_0$, describes a particle bound state propagating with a group velocity $v_g=(dE/dq)$, which has opposite sign for upper and lower minibands. In particular, one has $v_g>0$ for a wave packet belonging to the upper miniband provided that $-\pi/2<q_0<0$.
A pseudo-relativistic dynamics is obtained at the boundary of the Brillouin zone, where the dispersion relations of the two minibands Eq.(12)  can be approximated by the hyperbolic positive- and negative-energy branches of a one-dimensional (spinless) Dirac particle with an effective mass defined by the superlattice detuning parameter. In physical space, one should consider a broad wave packet with mean momentum close to the Bragg wave number. In this regime the discrete equations (8) can be transformed, by continuation of the variables, into the massive Dirac equation for a one-dimensional (spinless) particle. The mathematical derivation of the Dirac equation from the discrete tight-binding equations (8) can be found, for instance, in Refs.\cite{Cannata,LonghiOL}, and we refer the reader to such works for technical details.\\
Let us now consider the scattering problem from the potential step. According to Eqs.(8) and (11), the reflection of a two-particle bound state from the potential step (2) is formally analogous to the one-dimensional scattering problem of a single-particle in a binary superlattice by the potential step (11). As shown in Refs.\cite{K12,K19}, in this case partial transmission of a wave packet across the potential step can be observed  as a result of an {\it interband} tunneling process, which resembles KT of a massive relativistic Dirac particle \cite{note3}. Analytical calculations of the transmission coefficient of KT in this setting have been derived in Ref.\cite{K12} and compared with the expression of the transmission coefficient for a massive Dirac particle in the continuous limit of the discrete equations (8) \cite{notapalle}. Unlike KT for a massless Dirac particle, where complete transmission is achieved \cite{note}, for the case of a massive Dirac particle (which is actually the original tunneling problem studied by Klein and Sauter \cite{Klein,Calogero,Sauter}) the transmission is only fractional and related to the so-called kinematical factor  (see, for instance, Eqs.(2) and (3) of Ref.\cite{Calogero}).  
A physical picture of the two-particle interband tunneling process, and its connection to KT of a massive Dirac particle, is shown in Figs.2(b-d). The figures depict the space-dependent energy band diagram of the two-particle bound state and the interband tunneling process of a wave packet across the step that can be observed at large enough potential steps.  The two minibands depicted in the figure, and describing the states of a two-particle bosonic molecule in the original Hubbard model,  are analogous to the conduction (upper miniband) and valence (lower mniband) bands of a massive Dirac fermion in single-layer graphene (see, for example, \cite{cazzooo}).  As is well-known,  KT  is expected to be observed for an enough large  potential height such that  energy states of the conduction band (for $l<0$) are set in resonance with energy states of the  valence band (for $l>0$). Since we are dealing with a massive Dirac particle (i.e. the  dispersion curves of the two minibands are locally parabolic and separated by an energy gap, rather than being linear with no gap), wave packet transmission is only fractional and not complete. Indeed, let us consider a  wave packet belonging to the upper (conduction) miniband, with carrier wave number $q_0<0$ and energy $E_0=E_+(q_0)$, forward propagating along the lattice ($v_g>0$) and scattered off by the potential step at $l=0$. If the potential height $2\Delta$ is sufficiently low, the two-particle bound state undergoes  under-barrier tunneling (like for the single-particle problem), with the wave packet partially transmitted and partially reflected from the barrier; see Fig.2(b). As the barrier height is increased,  like for the single-particle case the potential step becomes impenetrable, and the two particles are fully reflected from the barrier; see Fig.2(c). As the barrier height is further increased, the lower miniband at $l>0$ becomes energetically overlapped with the upper miniband at $l<0$, and thus the two particles can now partially cross the region $l>0$ via an interband tunneling process; see Fig.2(d).  The transmitted wave packet, belonging to the lower miniband, has a carrier wave number $q_1>0$ which is obtained from the energy conservation relation $E_+(q_0)=E_-(q_1)+2 \Delta$, whereas the reflected wave packet, belonging to the upper miniband, has a carrier wave number opposite to the one of the incident wave packet, namely $-q_0$. The  transmission coefficient can be determined from the continuity condition of the wave function at the interface $l=0$, and its explicit form is given by Eq.(7) of Ref.\cite{K12}. Here we just mention that  the condition for the observation of  interband tunneling is that the energy $E_0$ of the incoming wave packet falls inside the lower miniband of the superlattice at $l>0$, i.e. 
\begin{equation}
 \Delta >  \frac{E_0-V}{2} \;\;\;  {\rm and} \;\;\; \Delta<\frac{E_0-V+W}{2}.
\end{equation}
Examples of KT for a bound particle wave packet will be presented in Sec.4.

\begin{figure*}
\resizebox{0.85\textwidth}{!}{
  \includegraphics{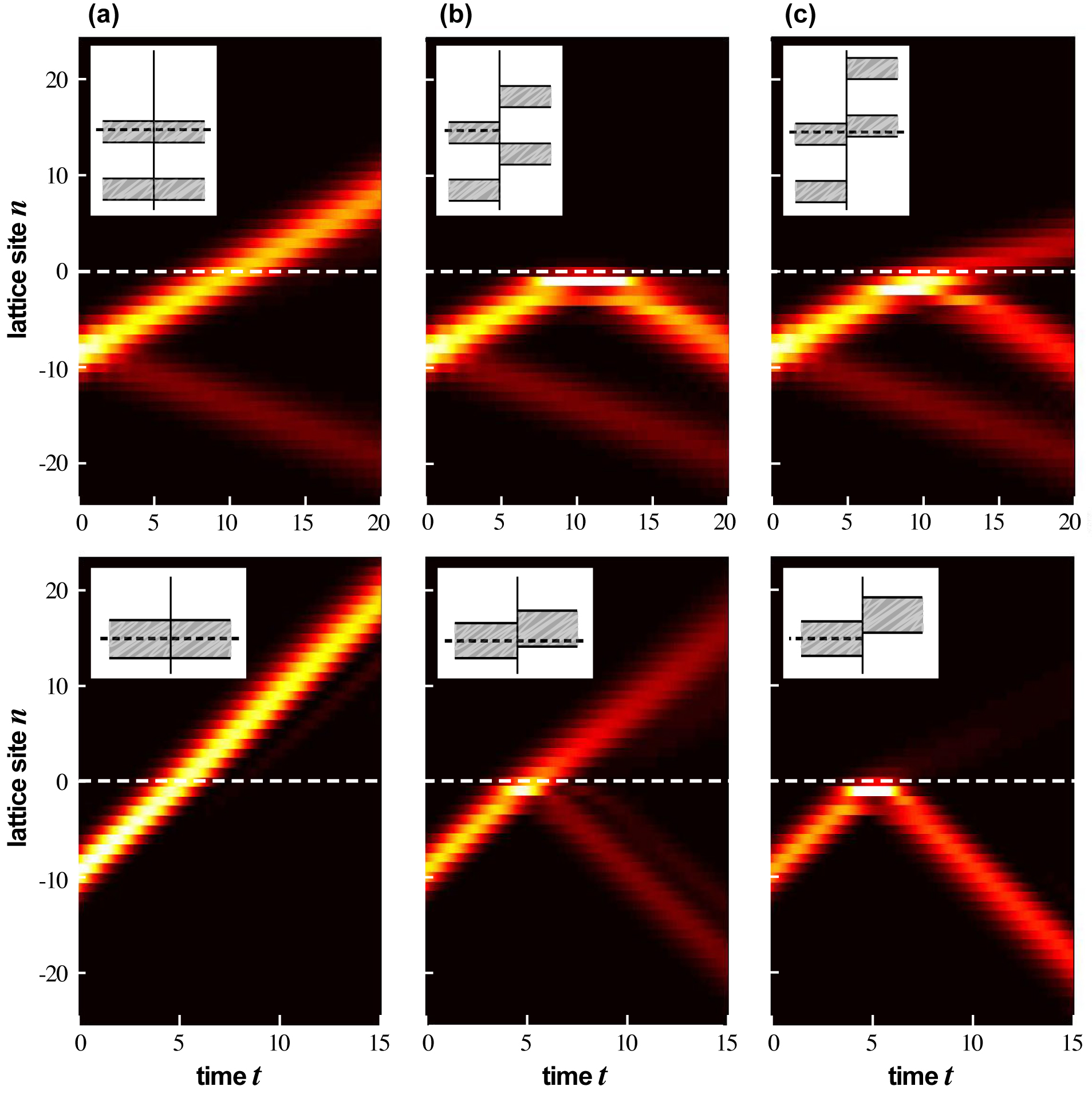}}
\caption{Scattering of a two-particle bound state wave packet from a potential step in the framework of the extended Bose-Hubbard model.  The upper panels show the numerically-computed evolution of the particle density function $P_n(t)$ [defined by Eq.(31)] for increasing values of the barrier height $\Delta$ and for $J=1$, $U=11$ and $V=8$. In (a) $\Delta=0$, in (b) $\Delta=1.5$, and in (c) $\Delta=2.6$. The horizontal dashed lines indicate the position of the potential step. The insets in the figures schematically show the space-dependent band diagrams of the two-particle bound state minibands in the three cases. In (c) KT is observed as an interband tunneling process. The lower panels show, for comparison, the scattering of a single-particle wave packet from the potential step [evolution of $|c_n(t)|^2$].}
\end{figure*}

  \section{Klein tunneling of the two-particle bound state in the ac-driven Bose-Hubbard model}
In the previous section we have shown that one-dimensional KT of a two-particle bound state, scattered off by a potential step, can be observed provided that the two particles show strong nearest-neighbor site interaction, in addition to onsite interaction. In this section we show that a similar result can be obtained even in the absence of nearest-neighbor particle interaction, provided that an external high-frequency ac driving force $F(t)$ is applied. To this aim, let us consider a standard Bose-Hubbard model describing the dynamics of on-site interacting bosons in the presence of an external driving field and of a potential step. The Hamiltonian of the system now reads 
\begin{eqnarray}
\hat{H} & = & - J \sum_{l} \hat{a}^{\dag}_{l} \left( \hat{a}_{l-1}+\hat{a}_{l+1}  \right) +\frac{U}{2} \sum_{l} \hat{n}_{l} (\hat{n}_{l}-1) \nonumber \\
&+ & \sum_l \epsilon_l \hat{n}_{l} + F(t) \sum_l l  \hat{n}_{l}.
\end{eqnarray}
As compared to the EHM of Sec.2 [see Eq.(1)], the nearest-neighbor interaction term $V$ is now absent in the Hamiltonian, however an external driving force $F(t)$ has been added. The external ac force can be introduced, for example, by periodically-shaking the optical lattice, as discussed in many works (see, for instance, \cite{Ari1,Ari2,Ari3} and references therein).
Let us focus our attention to the $N=2$ particle sector of Fock space, and let us expand the state vector $| \psi(t) \rangle$ of the system according to Eq.(4). The evolution equations of the amplitude probabilities $c_{n,m}(t)$ now read
 \begin{eqnarray}
 i \frac{dc_{n,m}}{dt}  & = & - J  \left( c_{n+1,m}+c_{n-1,m}+c_{n,m-1}
  + c_{n,m+1} \right)  \nonumber \\
  & + & \left[ U \delta_{n,m} +\epsilon_n+\epsilon_m+  (n+m) F(t) \right] c_{n,m}.
   \end{eqnarray}
 We consider a sinusoidal force at frequency $\omega$ and amplitude $F_0$
\begin{equation}
F(t)=F_0 \cos (\omega t)
\end{equation}
and assume the high-frequency and strong-interacting regimes, defined by 
\begin{equation}
 \frac{J}{ \omega} \equiv \alpha \ll 1, \;  \;  \frac{U}{\omega} \sim 1 , \; \; \frac{\Delta}{ \omega} \sim O(\alpha).
 \end{equation}
 In addition, we assume that the resonance condition $U \simeq M \omega$ is satisfied, 
 where $M$ is a non-vanishing integer number (typically $M=1$ or $M=2$). The detuning parameter
 \begin{equation}
 2 \sigma=U-M \omega
 \end{equation}
 from exact resonance is assumed to be small, such that $ \sigma / U \sim  O(\alpha)$.  To capture the dynamics of the two particles, it is worth introducing the new amplitudes
\begin{equation}
a_{n,m}(t)= c_{n,m}(t) \exp \left[ iM \omega \delta_{n,m}t+i(n+m) \Phi(t) \right]
\end{equation}
where we have set
\begin{equation}
\Phi(t)= \int_0^t dt' F(t')= \frac{F_0}{\omega} \sin (\omega t).
\end{equation}
 In terms of the new amplitudes $a_{n,m}$, the coupled equations (17) read
 \begin{eqnarray}
 i \frac{da_{n,m}}{dt} &= &  -J \left \{ a_{n+1,m} \exp \left[ i M \omega (\delta_{n,m}-\delta_{n+1,m}) t -i \Phi(t) \right] + \right. \nonumber \\
 & + & a_{n-1,m} \exp \left[ i M \omega (\delta_{n,m}-\delta_{n-1,m}) t +i \Phi(t) \right] +  \nonumber \\
 & + & a_{n,m+1} \exp \left[ i M \omega (\delta_{n,m}-\delta_{n,m+1}) t -i \Phi(t) \right] +   \\
 & + & \left. a_{n,m-1} \exp \left[ iM \omega (\delta_{n,m}-\delta_{n,m-1}) t +i \Phi(t) \right]  \right\} \nonumber \\
 & + & 2 \sigma \delta_{n,m} a_{n,m}+(\epsilon_n+\epsilon_m) a_{n,m} \nonumber 
 \end{eqnarray}
 In the high-frequency limit and assuming the scaling defined by Eqs.(19), at leading-order in the smallness parameter $\alpha$ the two-particle dynamics is described by neglecting the rapidly-oscillating terms in Eqs.(23) (see, for instance, \cite{uffi10,DellaValle}).  Application of the rotating-wave approximation to Eqs.(23) then leads to the following effective (averaged) set of coupled equations
 \begin{eqnarray}
 i \frac{da_{n,n}}{dt} & = & -2 J [J_M(\Gamma) a_{n,n+1}+J_{-M}(\Gamma) a_{n-1,n}] \nonumber \\
 & + & 2(\sigma + \epsilon_n) a_{n,n} \\
  i \frac{da_{n,n+1}}{dt} & = & -J [J_{-M} (\Gamma) a_{n+1,n+1}+J_M(\Gamma)a_{n,n} \nonumber \\ & + & J_0(\Gamma) a_{n-1,n+1}+J_0(\Gamma)a_{n,n+2}] \nonumber \\
  & + & (\epsilon_n+\epsilon_{n+1})a_{n,n+1} \\
   i \frac{da_{n,m}}{dt} & = & -J J_0(\Gamma) (a_{n+1,m}+a_{n-1,m}+a_{n,m+1}+a_{n,m-1}) \nonumber \\
   & + & (\epsilon_n+\epsilon_m)a_{n,m} \;\;\;\;\;\;\ (m>n+1)
 \end{eqnarray}
 where we have set
 \begin{equation}
 \Gamma = \frac{F_0}{\omega}
 \end{equation}
 and where $J_l$ is the Bessel function of first kind and of order $l$.  If the driving parameter $\Gamma$ is chosen such that $J_0(\Gamma)=0$ (for example at $\Gamma=2.405$), from Eqs.(24-26) it follows that the dynamics of amplitudes $a_{n,n}$ and $a_{n,n+1}$, governed by Eqs.(24) and (25),  decouples from the other amplitudes $a_{n,m}$ with $ m \geq n+2$. Note that in this regime the hopping of two uncorrelated particles on the lattice is suppressed, according to Eq.(26) (coherent destruction of tunneling \cite{Hanggi}). Hence, as opposed to the static Hubbard model considered in Sec.2.3, in the temporally-modulated Hubbard model hopping of a single boson on the lattice is suppressed, and thus it can not tunnel the barrier step. However, this is not the case for a bound particle state, for which hopping is not suppressed and thus tunneling across the potential step can be observed. The hopping motion of the two-particle bound state can be at best captured by introduction of the amplitudes
\begin{eqnarray}
f_{2n} & = & a_{n,n} \exp(-iM \pi n+i \sigma t) \\
 f_{2n+1} & = & \sqrt{2} a_{n,n+1} \exp(-iM \pi n+i \sigma t)
\end{eqnarray} 
 Taking into account that $J_{-M}(\Gamma)=(-1)^M J_M(\Gamma)$, substitution of Eqs.(28) and (29) into Eqs.(24) and (25) yields
 \begin{equation}
 i \frac{df_n}{dt}=-\sqrt{2} J_{eff} (f_{n+1}+f_{n-1})+(-1)^n \sigma f_n+\delta_n f_n
 \end{equation}
 where $\delta_n$ is defined by Eq.(10) and where we have set $J_{eff}=J J_M(\Gamma)$. Equation (30), which is the main result of this section, shows that the hopping dynamics of a two-particle bound state in the ac-driven Hubbard model, under the resonance condition (20) and provided that 
 $J_0(F_0 / \omega)=0$, is analogous to that of a two-particle bound state in the EHM, presented in Sec.2.2, where the difference $2 \sigma=(U-V)$ between on-site and nearest-neighbor site particle interaction energies is determined by the detuning of the driving quanta $M \omega$ from $U$ [see Eq.(20)] and the hopping rate $J$ is replaced by an effective hopping rate $J_{eff}= J J_M(\Gamma)$. Therefore, the main effect of the driving force, with appropriate detuning and amplitude, is to introduce a fictitious long-range (second-order) interaction in the original Hubbard model with on-site interaction solely, making it possible the observation of correlation-induced KT as discussed in Sec.2.3. A simple physical picture of the effective two-particle hopping dynamics in the presence of the high-frequency driving force, as described by Eq. (30), is the following one.  In the absence of the external driving force,  two bosons initially placed on the same lattice site form a stable bound state, dissociation being forbidden for energy constraint. For $J \ll U$, the hopping of the bound particle state on the lattice is very slow (it is a second order process), and can be neglected. When the external force is switched on, single-particle tunneling from the state $c_{n,n}$ (the two bosons occupy the same lattice site) to the state $c_{n,n \pm 1}$ (the two bosons occupy nearest-neighbor lattice sites) is allowed, energy conservation being now ensured by  $M$ quanta of external driving field ($U \sim M \omega$). In this way, the two-particle bound state $c_{n,n}$ can partially dissociate (into the state $c_{n,n \pm 1}$) and then recombine again in the nearest lattice site (the state $c_{n-1,n-1}$ or $c_{n+1,n+1}$), leading to an effective two-particle hopping motion on the lattice. If the resonance condition $U=M \omega$ is not strictly satisfied, the detuning $ 2 \sigma=U-M \omega$ introduces a residual energy mismatch $2 \sigma$ of the states, which mimics an effective unbalanced long-range (nearest-neighbor) interaction.

\begin{figure*}
\resizebox{0.85\textwidth}{!}{
  \includegraphics{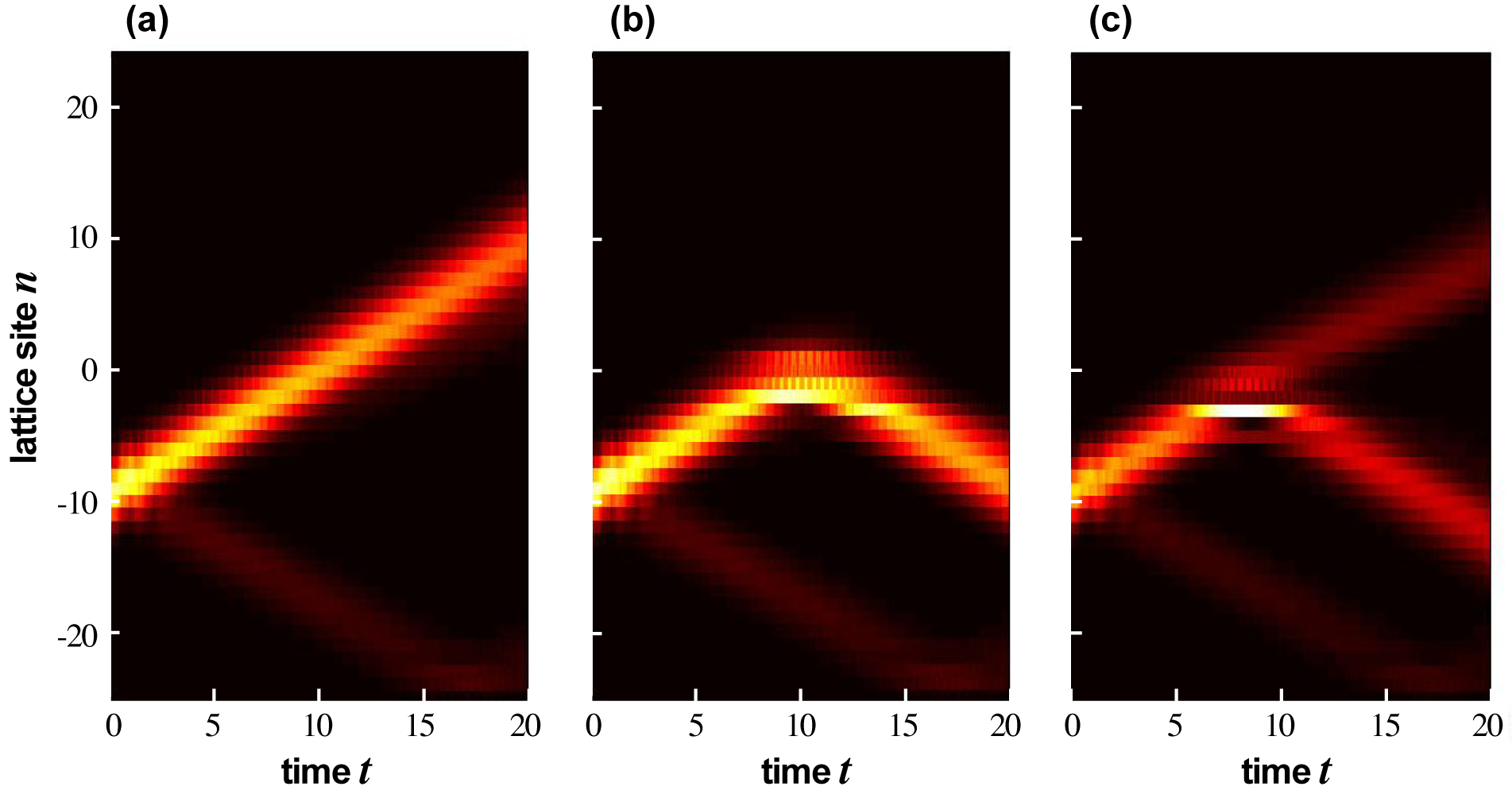}}
\caption{Scattering of a two-particle bound state wave packet from a potential step in the framework of the ac-driven Bose-Hubbard model.  The three panels show the numerically-computed evolution of the particle density function $P_n(t)$ [defined by Eq.(31)] for increasing values of the barrier height $\Delta$ and for $J=1.92$, $U=19$,
 $ \omega=16$, and $\Gamma=F_0 / \omega=2.405$. In (a) $\Delta=0$, in (b) $\Delta=1.5$, and in (c) $\Delta=2.6$.}
\end{figure*}

 \section{Numerical simulations}
 
To check the predictions of the theoretical analysis and the onset of KT for a bound particle state, investigated in Secs. II and III, we have numerically simulated the two-particle dynamics for the EHM and the ac-driven Bose-Hubbard model using an accurate variable-step fourth-order Runge-Kutta method, without any approximation. A wide lattice, comprising typically 50 sites, has been assumed to avoid finite boundary effects.\par
In a first set of simulations, we numerically solved the EHM (1) in the two-particle sector of Fock space to reveal the appearance of KT for the two-particle bound state. As an example, in Fig.3  we show a typical scattering  scenario for a two-particle bound state wave packet (upper panels), and corresponding scattering of a single-particle wave packet (lower panels),  for parameter values $J=1$, $U=11$, $V=8$ and for increasing values of the potential step height $\Delta$. For such parameter values, the band gap and width of the two bound particle state minibands of Fig.2(a)  are $2 \sigma=U-V=3$ and $W \simeq 1.7$, respectively. To simulate the scattering of a two-particle bound-state, Eqs.(5) have been numerically integrated with the initial condition $c_{n,m}(0) \propto \delta_{n,m} \exp[-(n+n_0)^2/w^2] \exp(2 iq_0n)$ and with $q_0=\pi/4$, $w=3$ and $n_0=8 $.  For the single particle problem, we assumed as an initial condition $c_{n} (0) \propto  \exp[-(n+n_0)^2/w^2] \exp(iq_0n)$ with $q_0=\pi/2$, $w=3$ and $n_0=9$. 
The upper panels in Figs. 3(a), (b) and (c) show the evolution of the particle density function
\begin{equation}
P_n(t)=\frac{1}{2} \langle \psi(t) | \hat{a}^{\dag}_n \hat{a}_n | \psi(t) \rangle
\end{equation}
 along the lattice for $\Delta=0$, $\Delta=1.5$, and $\Delta=2.6$. For the two-particle state, such barrier heights correspond to the absence of the barrier and to the cases (c) and (d) of Fig.2, respectively (see the insets in Fig.3).  For the single-particle problem, the three values of barrier height correspond to the absence of the barrier and to the cases (b) and (c) of Fig.1, respectively. For the two-particle problem, according to Eq.(6) the initial condition $c_{n,m}(0) \propto \delta_{n,m} \exp[-(n+n_0)^2/w^2] \exp(2iq_0n)$ corresponds to the excitation of the even-sites of the equivalent superlattice problem [Eq.(8)].  Such an initial condition mainly excites the upper miniband of the superlattice with spectral components centered at $q_0=\pi/4$, however a non-negligible superposition of Bloch modes belonging to the lower miniband also occurs. As a consequence, the initial two-particle wave packet splits into two wave packets forward and backward propagating, as clearly shown in the upper panels of Figs.3. The forward-propagating wave packet, belonging to the upper miniband, is then scattered off by the potential step, as discussed in Sec.2.3 [see also Figs.2(b-d)]. In the absence of the potential step,  the wave packet propagates straight away [see Fig.3(a)]. As the potential step is increased, the wave packet is partially transmitted and reflected by the step, till the barrier height gets larger than the energy of the wave packet.  In this regime the wave packet is fully reflected from the potential barrier, as clearly shown in the upper panel of Fig.3(b). As the barrier height is further increased, the lower miniband at $n>0$ gets overlapped with the upper miniband at $n<0$, and therefore KT is observed, as shown in Fig.3(c).  This behavior is not observed for a single particle, as shown in the lower plots of Fig.3. The numerical results corroborate the theoretical predictions of Sec.2 and clearly show that KT is a signature of particle correlation. Note that, since we are dealing with KT of a massive particle, wave packet transmission is not complete, and about $R \sim 62 \%$ of the wave packet is reflected at the interface. Such a value turns out to be in good agreement  with the theoretical value $R=1-T \simeq 0.598$ , computed from Eq.(7) of Ref.\cite{K12}. \par
 In a second set of simulations, we numerically solved the ac-driven Bose-Hubbard model (16) in the two-particle sector of Fock space to reveal the appearance of KT for the two-particle bound state even in the absence of nearest-neighbor particle interaction.  Parameter values used in the numerical simulations are $J=1.92$, $U=19$, $\omega=16$ and $\Gamma=2.405$, corresponding to the first ($M=1$) resonance condition [see Eq.(20)] with a detuning $ 2 \sigma=U-\omega=3$. Note that, for such parameter values the effective coupling $J_{eff}$ entering in the effective superlattice model of Eq.(30) is $J_{eff} = J J_1(\Gamma) \simeq 1$, so that the ac-driven Hubbard model basically maps the parameter values of the EHM shown in the simulations of Fig.3.  In Fig.4 we show the scattering  scenario for a two-particle bound state wave packet as obtained by solving Eqs.(17) for three values of the barrier height $\Delta=0$, $\Delta=1.5$ and $\Delta=2.6$. As an initial condition we assumed $c_{n,m}(0)\propto \delta_{n,m} \exp[-(n+n_0)^2/w^2] \exp(2iq_0n) (-1)^n$ with $q_0=\pi/4$, $w=3$ and $n_0=9$  \cite{note4}.  The scattering behavior of the wave packet is fully analogous to that observed in Fig.3 and is in agreement with the theoretical predictions presented in Sec.3.

 \section{Conclusions and discussion}
   
 In this work we studied the tunneling of two strongly-correlated particles across a potential step in the framework of the extended Bose-Hubbard  model and of the ac-driven 
 Bose-Hubbard model. The main result of the analysis is that, in the presence of nearest-neighbor particle interaction or, in the absence of this interaction, under a suitable ac-driving force, two strongly interacting particles forming  a bound state can undergo Klein tunneling across a high potential step, while a single particle does not. The reason thereof is that, contrary to a single-particle state which is described by a single tight-binding band, a two-particle bound state is described by two minibands, which are analogous to the positive- and negative-energy branches of the Dirac equation. When a potential step with appropriate height is applied, the upper and lower minibands can become overlapped, thus allowing for interband tunneling. This picture of KT shows that a two-particle bound state KT is analogous to KT of a single particle in a binary superlattice, investigated and experimentally observed in Refs.\cite{K12,K19}. However, the KT discussed in this work is physically very distinct from single-particle tunneling in a superlattice or from KT of single-particles in other physical systems (such as in graphene), because it is a clear signature of particle correlation.\par  
 As a final comment, we would like to briefly discuss possible model systems of the Bose-Hubbard Hamiltonian where correlation-induced KT could be observed. A first system is provided by cold atoms in optical lattices. The existence of two-atom bound states and correlated tunneling of pairs have been already observed in such systems \cite{uffi1,uffi2}. However, for the observation of KT of a bound particle state for ultracold atoms there are at least two issues that would deserve a further investigation. The first one is related to the system preparation of wo-particle bound state wave packets, which are  highly-excited states and require to properly drive the system of out equilibrium. 
Also, our analysis has been limited to consider tunneling of a single paired state, whereas with current experimental set-ups there are several bound pairs trapped in the optical lattice that undergo KT and that might interact. The second issue is the possibility to implement a sharp potential step. In fact, in a smooth potential barrier KT would be suppressed (see, for instance, \cite{K12}). 
Another experimentally-accessible and fully controllable model system capable of simulating the  two-particle sector of the extended Bose-Hubbard Hamiltonian is provided by light transport in square optical waveguide lattices  with diagonal defects \cite{Longhia,Longhib}. In this optical setting the temporal evolution of a two-particle system in Fock-space is mapped into spatial light propagation along a square waveguide lattice, with defects on the main and first two lateral diagonals that mimic on-site and nearest-neighbor particle interaction \cite{Longhia}. This optical system would enable to tune the difference $U-V$ in a  very simple way by changing the propagation constants of waveguides on the three diagonals of the lattice. Also excitation of the system with an elliptical Gaussian wave packet along the main diagonal, that  basically realizes the initial wave packet condition used in the simulations of Fig.3, should be feasible.\par
 To conclude, it is envisaged that our results could stimulate further theoretical and experimental studies on the simulation of relativistic quantum phenomena with correlated particles. For example, it is expected that a bound particle state freely hopping on the lattice should show Zitterbewegung, similarly to what happens to a relativistic freely moving particle \cite{Zitter}.


\begin{thebibliography}{}
\bibitem{note3}
 It should be noted that the potential barrier (11) experienced by the two-particle bound state hopping on the lattice is not strictly a sharp potential step, as for the single particle [compare Eqs.(11) and (2)]. However, as shown in Ref.\cite{K12}, KT is still observed for a rapid potential barrier. Only for a smooth (adiabatic) potential step KT is suppressed, as originally noticed by Sauter (see \cite{Sauter}).
\bibitem{note4}
The extra-phase term $(-1)^n$ in the initial condition is introduced because we are considering an odd resonance $M=1$, according to Eq.(28). Such term ensures that the wave packet mainly excited Bloch states belonging to the upper miniband of the equivalent superlattice. Such an extra-phase term would not be present for an even resonance. We also note that, for the ac-driven Hubbard model, 
the single particle would be frozen in the lattice owing to coherent destruction of tunneling, as discussed in Sec.3. Therefore, in Fig.4 we only show the scattering behavior of a two-particle state in the presence of particle interaction.
\end{thebibliography}
\end{document}